\documentclass[12pt]{article}%
\usepackage{amsfonts}
\usepackage{sw20elba}%
\usepackage{amsmath}%
\usepackage{amssymb}%
\usepackage{graphicx}
\providecommand{\U}[1]{\protect\rule{.1in}{.1in}}

\begin{document}

\title{Recent Studies in Superconductivity at Extreme Pressures}
\author{James S. Schilling and James J. Hamlin\\Department of Physics, Washington University\\CB 1105, One Brookings Dr, St. Louis, MO 63130, USA}
\date{October 12, 2007 }
\maketitle

\begin{abstract}
Studies of the effect of high pressure on superconductivity began in 1925 with
the seminal work of Sizoo and Onnes on Sn to 0.03 GPa and have continued up to
the present day to pressures in the 200 - 300 GPa range. Such enormous
pressures cause profound changes in all condensed matter properties, including
superconductivity. In high pressure experiments metallic elements, $T_{c}$
values have been elevated to temperatures as high as 20 K for Y at 115 GPa and
25 K for Ca at 160 GPa. These pressures are sufficient to turn many insulators
into metals and magnetics into superconductors. The changes will be
particularly dramatic when the pressure is sufficient to break up one or more
atomic shells. Recent results in superconductivity to Mbar pressures wll be
discussed which exemplify the progress made in this field over the past 82 years.

\end{abstract}

\pagebreak Superconductivity is a macroscopic quantum phenomenon which was
discovered by G. J. Holst and H. K. Onnes \cite{onnes1} in Leiden in 1911, but
not clearly understood until Bardeen, Cooper, and Schrieffer (BCS)
\cite{bardeen1} formulated their microscopic theory in 1957, exactly half a
century ago. In the opinion of one of the present authors (JSS), had
superconductivity not first been demonstrated in experiment, no theorist would
have ever predicted it: \ who could imagine that two electrons, in spite of
their Coulomb repulsion, might experience a net \textit{attractive}
interaction binding them together to form a bose particle? In a lecture in
1922 in honor of H. K. Onnes, Albert Einstein \cite{einstein1} considered
various scenarios but finally conceded that the theory of superconductivity
was in a sorry state with no end in sight. Three years later, and one year
before his death, Kamerlingh Onnes wrote the following introductory paragraph
in a paper of particular relevance to this conference \cite{onnes2}:
\ \textquotedblleft As no satisfactory theoretical explanation of the
supraconductive state of metals has been given yet, which might serve as a
guide for further investigations, it seems desirable to try, by changing the
external conditions, to discover the factors which play a roll in the
appearing of the phenomenon. These considerations led to the institution of an
inquiry into the influence which elastic deformation exerts on the appearing
of the supraconductive state. The results of this investigation are published
here.\textquotedblright\ In this paper the authors go on to present the
results of the first high-pressure experiments on any superconductor. Their
results on Sn are reproduced in Fig.~1 where the superconducting transition
temperature $T_{c}$ of Sn is seen to decrease by 5 mK under a hydrostatic
pressure of 300 bars. This paper is a model of careful experimentation and
should be read by every student engaged in high pressure research, whatever
the field of specialization.

The decrease in $T_{c}$ with pressure for Sn and In is, in fact, found for all
simple ($s,p$-electron) metal superconductors \cite{schilling1} and can be
easily accounted for within BCS theory which is based on an electron-phonon
pairing interaction:%
\begin{equation}
T_{c}\approx\sqrt{\frac{\kappa}{m}}e^{-\kappa/\eta},
\end{equation}
where $m$ is the cation mass, $\kappa$ is the lattice spring constant, and
$\eta$ is the Hopfield parameter \cite{hopfield1}, a purely electronic term.
In simple metals under pressure, $\kappa$ increases much \ more rapidly than
$\eta,$ so that the exponential factor, and thus $T_{c}$, decreases rapidly.
The decrease in $T_{c}$ with pressure in MgB$_{2},$ which \ with $T_{c}%
\simeq40$ K possesses the highest value of $T_{c}$ of any binary compound, can
also be accounted for assuming electron-phonon interactions \cite{deemyad1}.
Note that if the isotopic mass $m$ is varied in Eq.~1, instead of the
pressure, $T_{c}$ changes as $T_{c}\sim m^{-1/2},$ the classic isotope effect.

There are three primary ways in which high pressure experiments make important
contributions to the field of superconductivity:

\begin{enumerate}
\item Test theories of superconductivity and reveal important systematics (see above)

\item Improve the properties of known superconductors

\item Create new superconductors
\end{enumerate}

Besides testing theoretical models, high pressure experiments can also help
improve the superconducting properties by showing the experimentalist how to
push materials to higher values of $T_{c}.$ This is of great technological
relevance since no superconductor has yet been found which at ambient pressure
possesses a transition temperature even half that of room temperature, in
contrast to magnetic materials which order at much higher temperatures and
have had an enormous commercial impact. For this reason the pressure
dependence of $T_{c}$ is often one of the first experiments to be carried out
after a new superconductor is discovered: \ a large value of $dT_{c}/dP,$
\ either positive or negative, implies the material is likely capable of
higher transition temperatures at ambient pressure either through selective
doping or positive or negative chemical pressure. This strategy was applied by
Paul Chu's group with great success in the late 1980's. They determined that
the transition temperature of the oxide cuprate La-Ba-Cu-O increased from 32 K
to 40 K under only 1.4 GPa pressure \cite{chu1}; the replacement of the large
La cation with the smaller Y generated chemical pressure and resulted in the
discovery of the first superconductor Y-Ba-Cu-O with a transition temperature
($T_{c}\simeq92$ K) above the boiling point of liquid nitrogen \cite{chu2}.
Several years later, in a collaboration with members of the Carnegie Institute
of Washington, Chu's group was able to push the transition temperature of the
mercury compound Hg-1223 to the record-high value $T_{c}\approx160$ K
\cite{chu3}. These are excellent examples of how high pressures can be used to
improve the properties of superconducting materials which exemplify the
philosophy and tradition of the group of Bernd Matthias at the University of
California, San Diego which has been ably carried forth by many of his
students, including Paul Chu and Matthias' successor, Brian Maple.

High pressures can also be used to synthesize new superconductors;
simultaneous high temperature may be necessary \cite{HT}, but sometimes not.
In Fig.~2 is an updated and expanded version of the Periodic Table of
Superconductivity originally drawn up by Bernd Matthias. There are altogether
52 elemental superconductors, 22 of which are only known to superconduct if
sufficient pressure is applied. Of these 22, fully 12 were discovered by
J\"{o}rg Wittig, a student of Werner Buckel at the University of Karlsruhe,
Germany. Across the periodic table values of $T_{c}$ range from 0.325 mK for
Rh \cite{buchal} or 0.4 mK for Li \cite{tuor} (at 1 bar) to 20 K for Y (at 115
GPa) \cite{hamlin1} or 25 K for Ca (at 160 GPa) \cite{shimizu1}. This record
high value of $T_{c}$ for Ca was discovered by the group of Katsuya Shimizu, a
student, and the successor, of Katsuya Amaya at the University of Osaka,
Japan. Pressure-induced superconductivity is of great interest from a basic
physics point of view since in a single high pressure experiment both the
birth and the demise of the superconducting state can be witnessed. After a
brief discussion of relevant high-pressure technology, we will examine several
classes of pressure-induced transitions from the normal to the superconducting state.

High pressure technology has come a long way since the experiments of Sizoo
and Onnes in 1925 on Sn and In at 300 bar pressure \cite{onnes2}. In the early
1970's Webb, Gubser, and Towle developed a diamond-anvil cell (DAC) mated to
the mixing chamber of a dilution refrigerator with a SQUID pickup coil capable
of detecting minute magnetic flux changes \cite{webb1}. With this beautiful
device they tracked the decreasing $T_{c}$ of Al metal from 1.17 K to 80 mK at
6.1 GPa \cite{gubser1}. Using this DAC as a model, one of the present authors
(JSS) constructed a DAC suitable for higher pressures where superfluid He was
loaded into the gasket hole to serve as pressure medium \cite{schilling2}. The
opposing diamond anvils were then pushed into the gasket at 2 K using a He-gas
driven double membrane. Thus far pressures as high as 220 GPa have been
generated in this DAC at cryogenic temperatures, if no pressure medium is
used. Magnetic flux changes are detected by one of two matched counterwound
pickup coils connected either to a lock-in amplifier or a DC SQUID. Shimizu
\textit{et al.} \cite{shimizu3} utilize a DAC designed to generate Mbar
pressures at temperatures as low as 5 mK; this DAC made possible the discovery
of pressure-induced superconductivity in O$_{2}$ near 100 GPa \cite{shimizu4}.

An enhancement in the detection limit for superconductivity is provided by the
double modulation technique utilized by Timofeev \textit{et al.}
\cite{filipek} at the Carnegie Institute in Washington. Vohra and Weir have
developed so-called designer diamond anvils\ where up to 8 electrical leads or
pickup loops are grown into the diamond anvil itself \cite{vohra1}. These
anvils offer great promise for future quantitative electrical resistivity or
ac susceptibility measurements in the multi-Mbar pressure range.

Let us now return to the Periodic Table of Superconductivity in Fig.~2. An
examination of the maximum measured values of $T_{c}$ at ambient or high
pressure reveals that the lighter elements tend to do better. This is
particularly evident for the alkaline earths going from Ba (5 K) to Sr (7 K)
to Ca (25 K) and for the alkali metals from Cs (1.3 K) to Li \ (14 K). The
lightest element, of course, is hydrogen. Many years ago Ashcroft
\cite{ashcroft1} predicted that superconductivity near room temperature might
be in the offing if only enough pressure is applied to make it metallic. Alas,
the metallic state in hydrogen has not yet been sighted, except at very high
temperatures (3000 K) \cite{nellis1} where superconductivity has little
chance. Should metallic hydrogen prove to be fluid at low temperatures, then
two-component superconductivity from the electrons and protons has been
predicted with a myriad of fascinating ground states both with and without an
applied magnetic field \cite{babaev1}. Is it realistic to hope that metallic
hydrogen will be fluid at low temperatures in some pressure range? A hint at
an answer is given by the fact that the melting curve of Na drops
precipitously from 1000 K at 30 GPa to 300 K at 118 GPa \cite{greg}, a
significantly lower melting temperature than at ambient pressure!

As seen in Fig.~2, the Periodic Table of Superconductivity contains four
classes of elemental solids that refuse to superconduct at ambient pressure:
\ (i) the non-metals like H$_{2}$, O$_{2}$, Si, or Ge, (ii) elements like Fe,
Co, Ni, Mn, as well as most rare earths and many actinides which are magnetic,
the arch-enemy of superconductivity, (iii) the early, trivalent $d$-electron
elements Sc, Y, and Lu, and (iv) the monovalent alkali and noble metals,
excepting Li which barely makes it with $T_{c}\simeq0.4$ mK. The first class
is the easiest to understand: \ without conduction electrons,
superconductivity doesn't have a chance. How does one predict what pressure is
necessary to turn an insulator into a metal? With ample financial resources
one can carry out a full scale electronic structure calculation or, as a
cost-saving alternative, one can use the simple Goldhammer-Herzfeld criterion
\cite{goldhammer} for metallization which states that a given substance should
be metallic if the ratio of its molar refractivity to its molar volume is
equal to or greater than 1, i.e. $R/V\geq1.$ Using the equation of state, it
is thus trivial to estimate the metallization pressure if the initial value of
$R/V$ is known (the change in R with pressure can be neglected). $R$ is
obtained from the measured atomic polarizability $\alpha, $ where
$R=(4\pi\alpha/3)N_{A}$ and $N_{A}$ is Avagadro's number. In Fig.~3 the ratio
$R/V$ is plotted for all elements (upper figure) \cite{edwards1} and compared
to the actual electronic state of all elemental solids (lower figure). The
Goldhammer-Herzfeld criterion is seen to have an amazing predictive power.

The fact that diamond anvils are transparent to visible light allows optical
transmission and absorption experiments in a DAC to determine the band gap
$E_{g}$ of an insulating material as a function of pressure. If white light is
used, the light transmitted through the sample will appear yellow, orange, or
black if $E_{g}\approx2.2,$ 1.7, or 1.4 eV, respectively. Ashcroft
\cite{ashcroft2} has recently pointed out that in hydrogen-rich compounds,
such as LiBH$_{4}$ and LiAlH$_{4}$, the hydrogen may be subjected to lattice
pressure which could significantly reduce the external pressure required for
metallization.\ In Fig.~4 microscopic photographs of the light transmitted
through the two insulating compounds LiBH$_{4}$ and LiAlH$_{4}$ are shown for
a sequence of increasing pressures to $\sim$ 50 GPa. In LiBH$_{4}$ no
coloration of the white light is seen to 53 GPa, indicating that the band gap
remains greater than 2.2 eV throughout the experiment. In contrast, in
LiAlH$_{4}$ the transmitted light begins to change color at 30.8 GPa, becoming
progressively more red to 42 GPa; however, further loading to 75 GPa resulted
in no further change in color. The fact that upon unloading the color change
is not reversible gives evidence for an irreversible phase transition.

The second class of elemental solids that do not superconduct at ambient
pressure are those which are magnetic: the 3$d$ transition metals, the rare
earths (4$f$), and the actinides (5$f$). This is a vast playground for high
pressure research where the possibilities include both transitions from one
form of magnetism to another as well as from magnetism to superconductivity
\cite{schilling2}. For example, Shimizu \textit{et al.}
\cite{shimizu3,shimizu5} have shown that Fe loses it's ferromagnetism when it
transforms from the bcc to the hcp $\epsilon$-phase near 14 GPa, allowing
superconductivity to appear with $T_{c}$ reaching a maximum value of 2.1 K at
21 GPa. For lack of space we will not discuss pressure-induced transitions in
magnetic substances further.

The third class of non-superconducting elemental solids are the early,
trivalent $d$-electron elements Sc, Y, and Lu. Why is superconductivity
lacking here? The simple answer is that they don't have a sufficient number of
$d$ electrons in the conduction band. The more $d$ electrons a given
conduction band has, the higher the density of states and the more likely it
is to support superconductivity. Superconductivity in Sc \cite{wittig2}, Y
\cite{wittig1}, and Lu \cite{wittig3} was discovered in high pressure
experiments by Wittig and coworkers. This is not surprising since it is well
known that the application of pressure increases the number of $d$ electrons
in a band. This $s-d$ transfer is expected on very general grounds
\cite{mcmahan1}: the kinetic energy $\sim V^{-2/3}$ increases much faster
under pressure than the potential energy $\sim V^{-1/3}$ ; since $s$-orbitals
possess more radial nodes (higher kinetic energy) than $d$-orbitals, they
shift to higher energy much faster under pressure, thus causing $s$ electrons
to be dumped into the $d$ band. Hamlin \textit{et al.} have recently taken Y,
Sc, and Lu to pressures as high as 115 \cite{hamlin1}, 74 \cite{hamlin2}, and
174 GPa \cite{hamlin3}, respectively, and found that $T_{c}$ increases to
temperatures as high as 20 K, as seen in Fig.~5. The upward curvature in
$T_{c}(P)$ for Sc suggests that extending the experiments into the Mbar
pressure range may result in a significant enhancement in $T_{c}.$ Such
experiments on Sc are currently underway.

One nagging questions remains: \ why is the early, trivalent $d$-electron
metal La superconducting at ambient pressure, but Sc, Y, and Lu are not? We
will return to this question after considering the alkali metals.

Superconductivity in the alkali metals has recently been reviewed by one of
the present authors (JSS) \cite{schilling3}. The first alkali metal to exhibit
superconductivity was discovered by Wittig \cite{wittig1} in 1970 for Cs at
1.3 K under 12 GPa pressure. Electronic structure calculations \cite{grover1}
reveal that appreciable $s-d$ transfer (6$s$ to 5$d$) occurs in this pressure
range, i.e. Cs becomes a transition metal. The appearance of superconductivity
is thus not surprising. However, if pressure-induced $s-d$ transfer is
responsible for the superconductivity of Sc, Y, Lu, and now Cs, how can it be
that Li, where $s-d$ transfer plays no role, becomes superconducting under
pressure at relatively high temperatures (14 K)? The alkali metals are
believed to be model free electron systems with nearly spherical Fermi
surfaces. The particularly low electronic density of states of monovalent
free-electron systems, the alkali and noble metals, contributes strongly to
the lack of superconductivity at ambient pressure, with the exception of Li
where $T_{c}$ lies at extremely low temperatures (0.4 mK). In $s,p$-electron
system, like Pb, Sn, In, or Zn, which manage to become superconducting at
ambient pressure, $T_{c}$ invariably \textit{decreases} under pressure, as
pointed out above. It would, therefore, appear highly unlikely that pressure
would induce superconductivity in Li, Na or the noble metals where $s-d$
transfer is relatively unimportant.

The resolution of this apparent paradox was given by Neaton and Ashcroft who
calculated the electronic properties of Li \cite{neaton1} and Na
\cite{neaton2} at Mbar pressures and obtained for Li the remarkable
2$s$-electron density shown in Fig.~6. Not only is a clear tendency evident
that the Li cations pair up, but the 2$s$ electrons are seen to be
concentrated in interstitial regions rather than between the paired Li
cations. The electronic properties they obtain at Mbar pressures are highly
anomalous, the conduction bandwidth even \textit{narrowing} under compression,
a highly counterintuitive result in direct opposition to standard textbook
dogma. The electron-phonon interaction also increases strongly. The physical
picture they offer can be illustrated using Fig.~7 which depicts the ion cores
as well separated at ambient pressure but beginning to touch at Mbar
pressures. Since the conduction electrons must avoid the cation core region
because of Pauli principle and orthogonality constraints, the free volume
available to them outside the ion cores becomes quite small under Mbar
pressures, forcing these electrons into the interstitial sites and away from
the regions between the cation cores. The low symmetry of the interstitial
sites forces the conduction electrons to take on higher angular momentum
character, i.e. for $s$ electrons more $p$ or $d$ character, thus improving
the likelihood that superconductivity occurs. This picture is not restricted
to the alkali metals, but is quite general \cite{neaton2} and also applicable
to the transition metals, as we discuss below. In Li and Na, therefore, one
may speak of significant pressure-induced $s-p$ rather than $s-d$ transfer.

In a very recent paper Feng \textit{et al.} \cite{feng1} pose the interesting
quetion whether the anomalous properties predicted and found for Li and Ca
under pressure carry over to CaLi$_{2},$ the only known binary compound
containing these two elements. Detailed calculations by these authors lead
them to predict that \textquotedblleft the elevated density of states at the
Fermi level, coupled with the expected high dynamical scale of Li as well as
the possibility of favorable interlayer phonons, points to potential
superconductivity of CaLi$_{2}$ under pressure\textquotedblright. We have
carried out a search for superconductivity in CaLi$_{2}$ both at ambient
pressure, where no superconductivity was found above 1.10 K, and high pressure
\cite{matsuoka1}. The results of the latter electrical resistivity experiments
are shown in Fig.~8 where a superconducting transition is seen to appear below
2 K at 11 GPa. With a further increase in pressure, $T_{c}$ increases and
passes through a maximum at 13 K near 40 GPa.

The physical picture in Fig.~7, that the free volume available to the
conduction electrons is sharply reduced as pressure brings the ion cores
together, can also be applied to the early, trivalent $d$-electron metals Sc,
Y, La, and Lu. This was first pointed out by Johansson and Rosengren
\cite{johansson} who were able to analyze $T_{c}(P)$ data for La and La-Y
alloys in terms of the reduction in the free volume available to the
conduction electrons as pressure is applied. They used the ratio $r_{a}/r_{c}
$ of the Wigner-Seitz radius to that of the cation core as a measure of this
free volume. In Fig.~9 we plot $T_{c}$ versus ratio $r_{a}/r_{c}$ for Sc, Y,
La, and Lu, where we neglect any change in $r_{c}$ with pressure compared to
that in $r_{a}$, where $r_{a}\propto V^{1/3}.$ For superconductivity to set in
under pressure, evidently the ratio $r_{a}/r_{c}$ must decrease to the value
2.2 or smaller. The vertical arrows mark the value of $r_{a}/r_{c}$ for each
elemental metal at ambient pressure. Note that for these four elemental metals
the ambient pressure value of $r_{a}/r_{c}$ is smallest for La which is
consistent with the fact that La is already superconducting at ambient
pressure, but Sc, Y, and Lu are not.

The alkali metals are also included in Fig.~9. Li is seen to fit in quite well
with the trivalent $d$-electron systems. However K, Rb, and Cs's ambient
pressure value of $r_{a}/r_{c}$ would lead to the expectation that they would
superconduct at ambient pressure, which they don't. Perhaps our neglect of any
change in the core radius $r_{c}$ with pressure is oversimplified, as is our
attempt to compare the properties of monovalent and trivalent elemental solids.

If the electronic properties of conduction electrons in a solid become highly
anomalous as the ion cores approach each other under very high pressure, how
much greater the changes would be if sufficient pressure is applied to
actually begin to break open the individual atomic shells inside the ion core!
As each atomic shell is broken open and its electrons dumped into the
conduction band, all solid state electronic properties would fluctuate wildly.
Things finally settle down when at astronomic pressures all shells have been
broken open and we have a structureless Thomas-Fermi electron gas. Here it
would seem likely that superconductivity and magnetism would weaken and
ultimately disappear. This progression of events is mirrored schematically in
the changes in the superconducting transition temperature shown in Fig.~10
where $T_{c}$ is seen to fluctuate strongly with pressure. In our experiments
to only a few Mbars we are only probing a relatively modest fraction of this
mountain range. Science at high pressures has an exciting future to look
forward to!\vspace{0.2cm}

\noindent Acknowledgments. \ The authors would like to thank M. Debessai, W.
Bi, and N. Hillier for experimental assistance. Stimulating discussions with
N.W. Ashcroft are acknowledged. The authors are particularly grateful to Ms.
Dani\"{e}lle Duijn of the Kamerlingh Onnes Laboratory in Leiden for sending us
a reprint of Sizoo and Onnes' seminal publication from 1925. Thanks to A.
Hofmeister for allowing us to use the optical DAC made by R. B\"{o}hler. This
research was supported by the National Science Foundation through grant No. DMR-0404505.

\begin{center}
\bigskip{\LARGE Figure Captions}
\end{center}

\bigskip\ 

\noindent\textbf{Fig. 1. \ }Electrical resistance versus temperature (units of
$^{4}$He vapor pressure) under hydrostatic pressures of 4, 95, 193, and 300
bar from Ref.~\cite{onnes2}. $T_{c}$ decreases under pressure. Figure
reproduced with permission from the Kamerlingh Onnes Laboratory, Leiden, The
Netherlands.\bigskip

\noindent\textbf{Fig. 2. \ }Periodic Table of Superconductivity listing 30
elements which superconduct at ambient pressure (yellow) and 22 elements which
only superconduct under high pressure (green, bold black frame). For each
element the upper position gives the value of $T_{c}$(K) at ambient pressure;
middle position gives maximum value $T_{c}^{\max}$(K) in a high-pressure
experiment at $P$(GPa) (lower position). $T_{c}$ values from
Refs.~\cite{schilling1,tuor}.\bigskip

\noindent\textbf{Fig.\ 3.\ \ }(upper) Ratio of atomic molar refractivity
versus molar volume $R/V$ for elemental solids from Ref.~ \cite{edwards1}.
(lower) Corresponding periodic table where metals, semimetals/semiconductors,
and insulators are color coded.\bigskip

\noindent\textbf{Fig.\ 4.\ \ }Transmitted light micro-photographs of
LiBH$_{4}$ and LiAlH$_{4}$ at ambient temperature to 53 and 42 GPa,
respectively. Sample diameter is $\sim100$ $\mu$m. Color gives estimate of
band gap: \ $E_{g}\approx$ 2.2 eV (yellow), 1.7 eV (orange), 1.4 eV
(black).\bigskip

\noindent\textbf{Fig. 5. \ }Superconducting transition temperature versus
pressure for Sc \cite{hamlin2}, Y \cite{hamlin1}, and Lu \cite{hamlin3}%
.\bigskip

\noindent\textbf{Fig. 6. \ }Relative charge density of Li's 2$s$ conduction
electrons at 100 GPa pressure. The Li$^{1+}$ ions are centered in the red
regions. Reprinted by permission from Macmillan Publishers Ltd: \ [NATURE]
Neaton J B and Ashcroft N W 1999 \textit{Nature} \textbf{400} 1141, copyright
(1999).\bigskip

\noindent\textbf{Fig. 7. \ }Scematic representation of cation electron cores
at (left) ambient pressure and (right) extreme pressure. $r_{c}$ and $r_{a}$
define the core and Wigner-Seitz radii, respectively.\bigskip

\noindent\textbf{Fig. 8. \ }Resistance versus temperature for annealed
CaLi$_{2}$ at 13.4, 20, 22, and 24 GPa. Inset gives expanded view of
superconducting transition at 24 GPa for dc magnetic fields of 500 Oe (2 runs)
and 0 Oe (3 runs), where colors distinguish runs.\bigskip

\noindent\textbf{Fig. 9. \ }Pressure-induced superconducting transition
temperature versus ratio $r_{a}/r_{c}$ for the trivalent $d$-electron metals
Sc, Y, La, Lu and the alkali metals Li and Cs. Vertical arrows mark values of
$r_{a}/r_{c}$ at ambient pressure.\bigskip

\noindent\textbf{Fig. 10. \ }Schematic representation of the superconducting
transition temperature versus (astronomic) pressures showing the drastic
changes which occur when the atomic shell structure is progressively broken up.

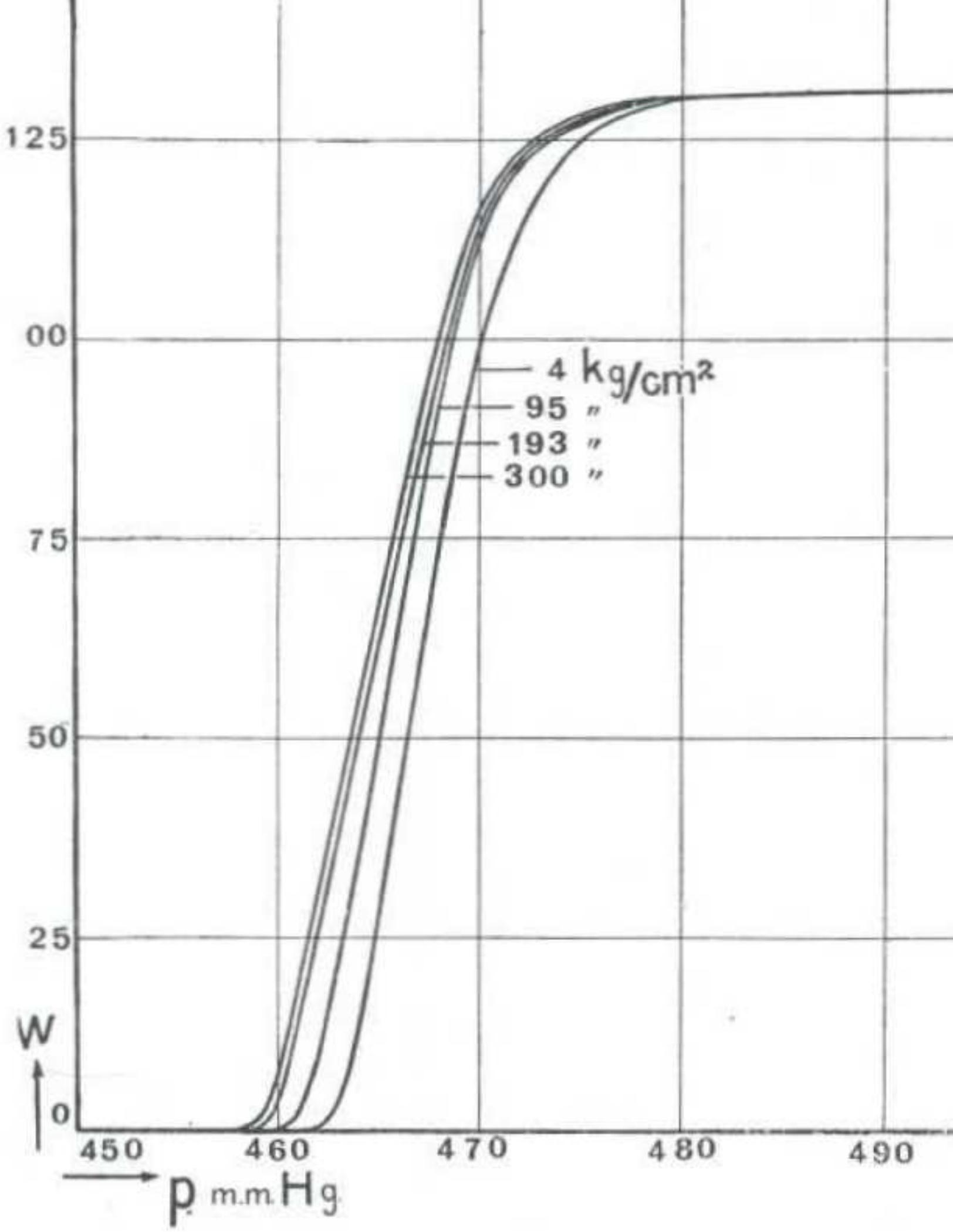
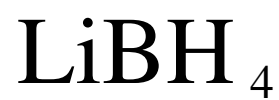
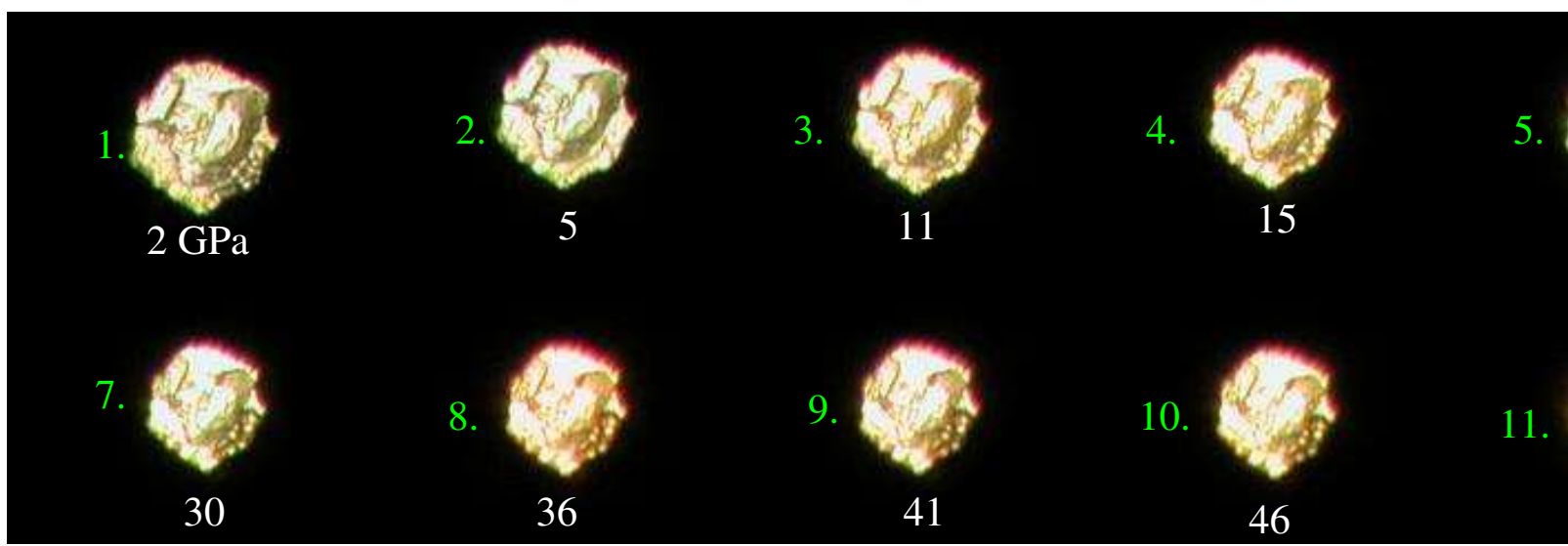
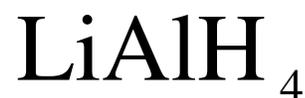
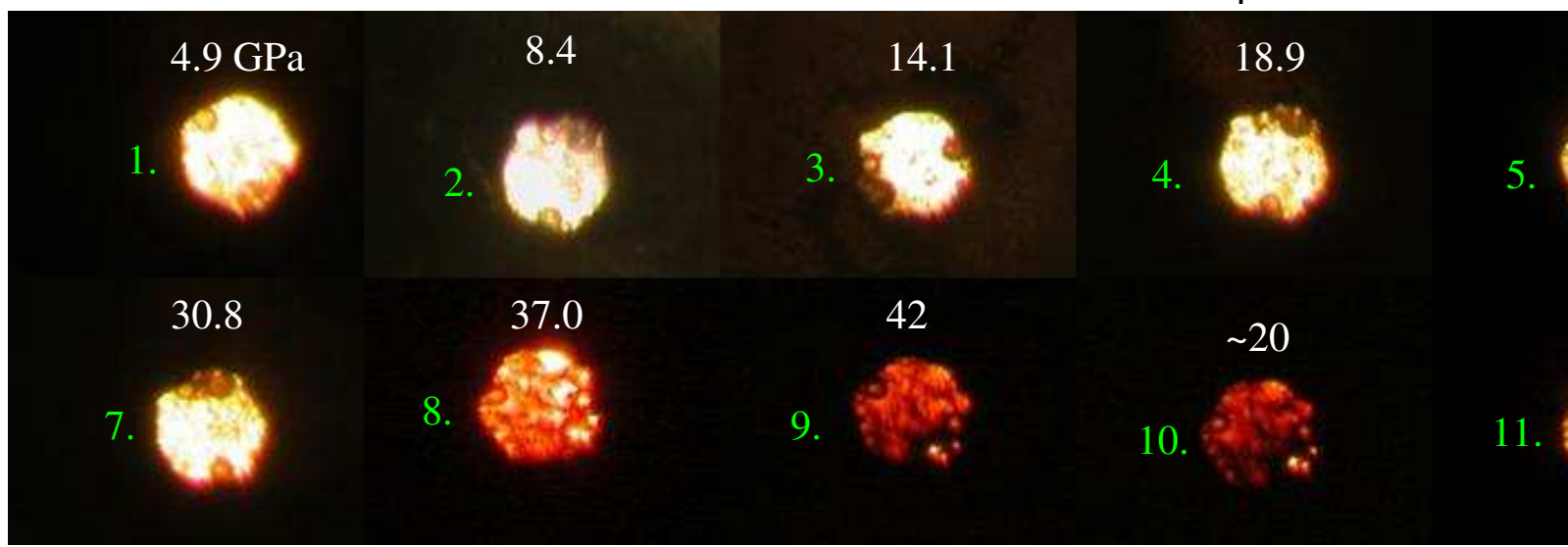
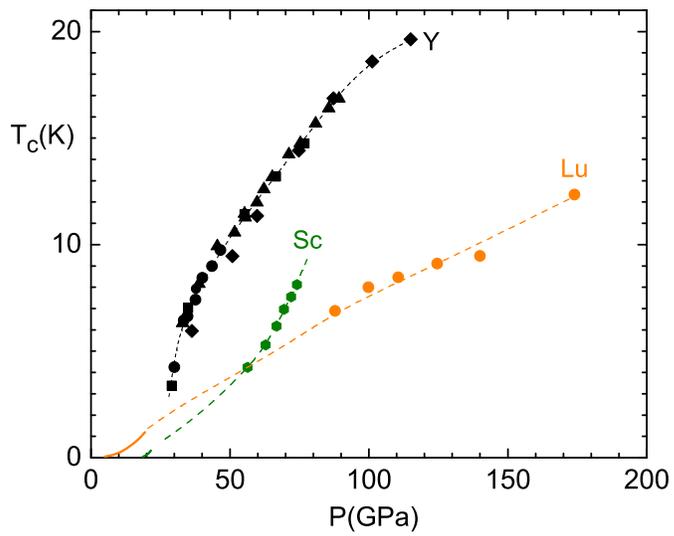
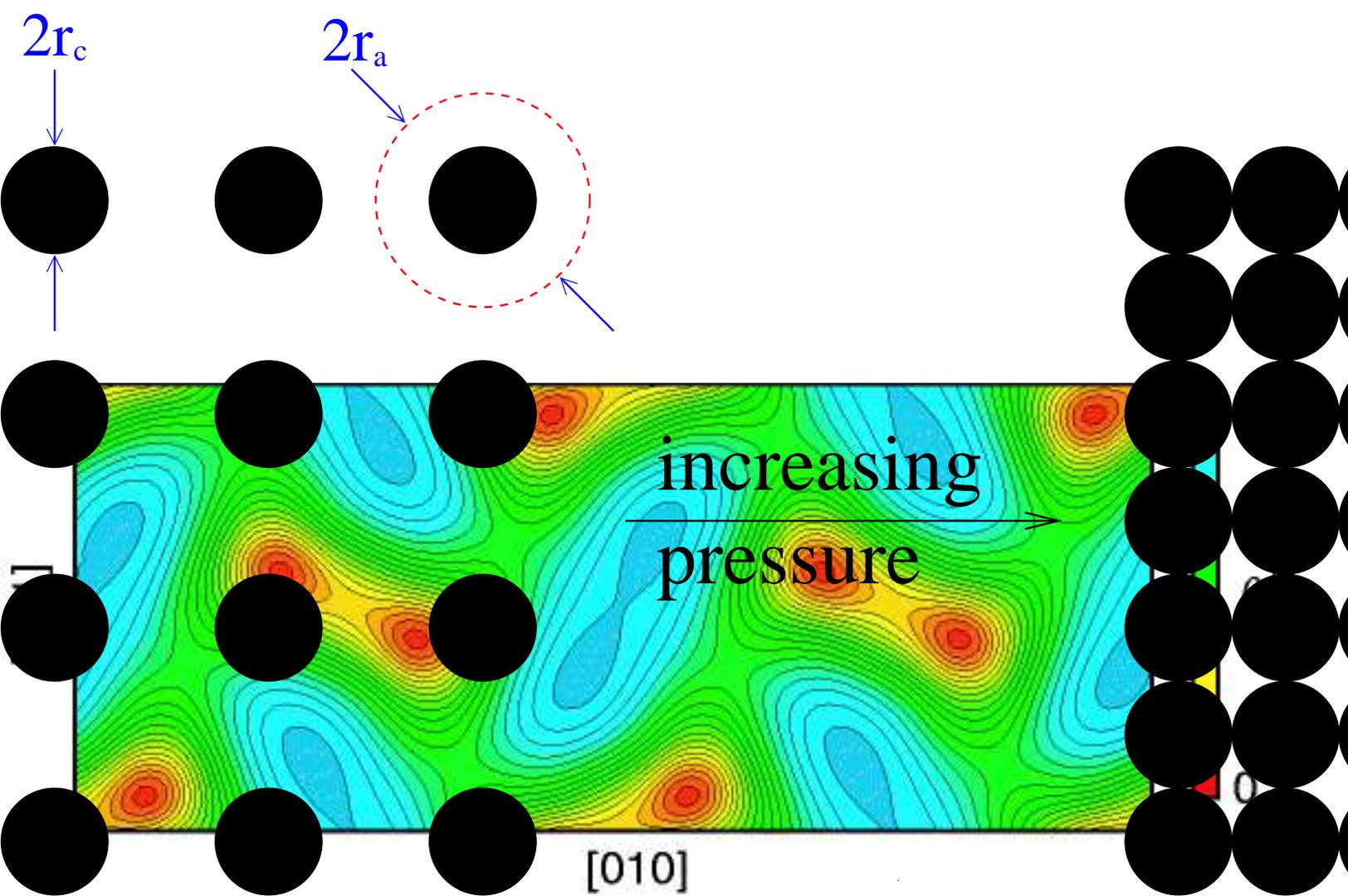
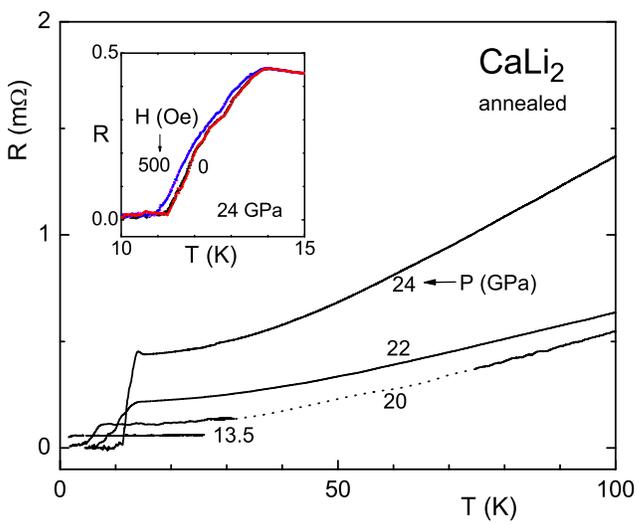
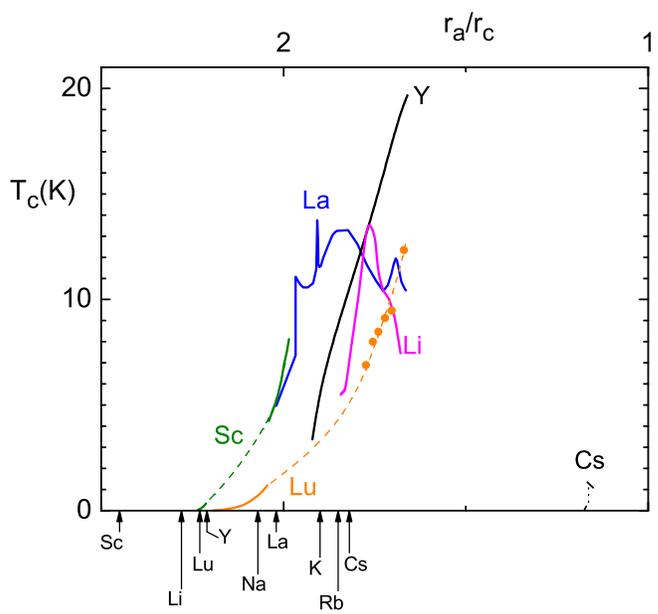
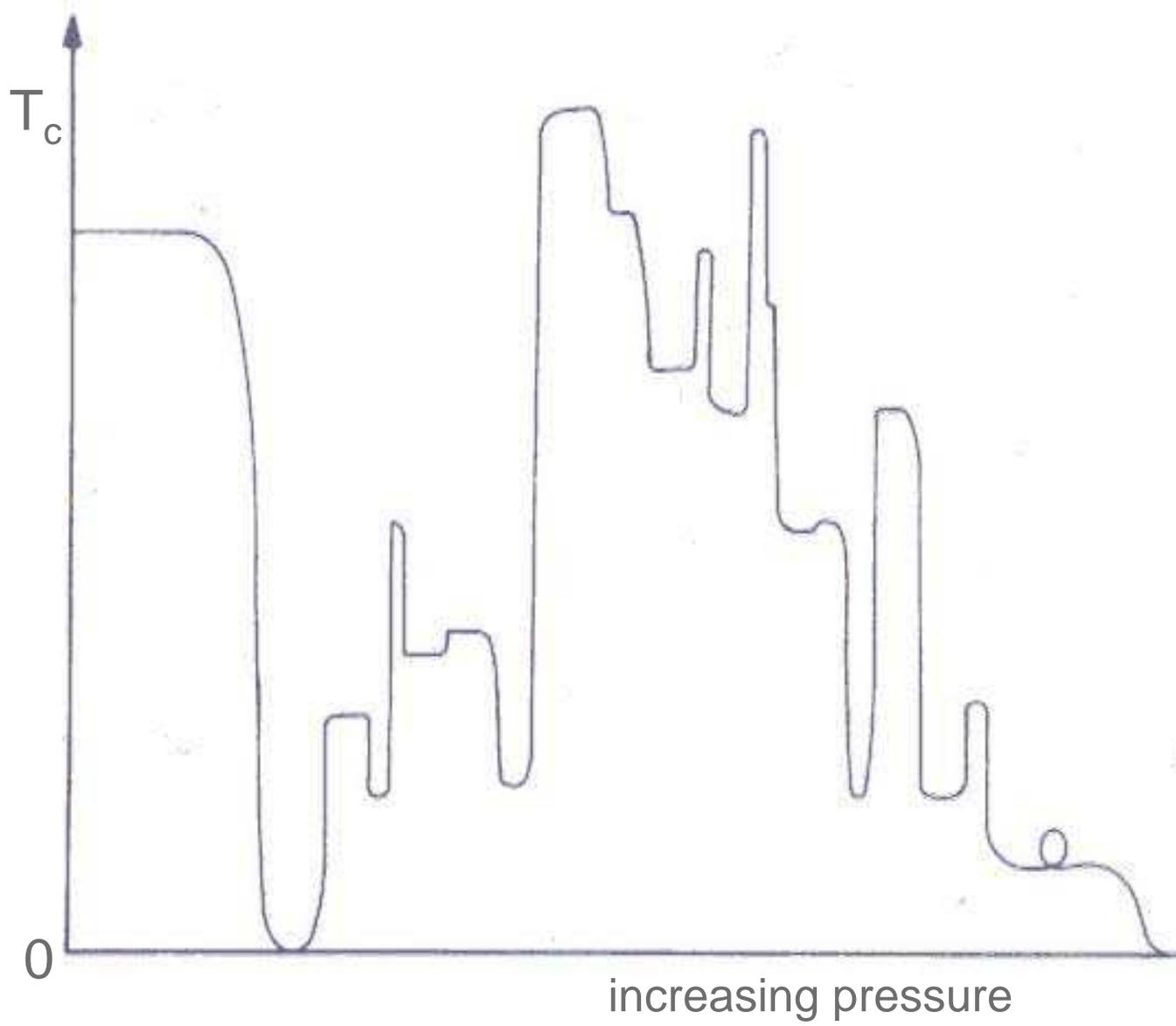
\end{document}